\documentclass[conference]{IEEEtran}
\IEEEoverridecommandlockouts
\usepackage{cite}
\ifCLASSINFOpdf
  \usepackage[pdftex]{graphicx}
  \graphicspath{{../pdf/}{../jpeg/}}
  \DeclareGraphicsExtensions{.pdf,.jpeg,.png}
\else
  \usepackage[dvips]{graphicx}
  \graphicspath{{../eps/}}
  \DeclareGraphicsExtensions{.eps}
\fi
\usepackage{amsmath}
\usepackage{amsfonts}
\usepackage[dvipsnames]{xcolor}
\usepackage{amssymb}
\usepackage{array}
\usepackage{multirow}
\usepackage{longtable}
\usepackage[english]{babel}
\usepackage{amsthm}
\usepackage{scalerel}
\usepackage{hyperref}

\usepackage{textcomp}
\usepackage{url}
\usepackage{xcolor}

\allowdisplaybreaks

\usepackage[
linesnumbered,ruled,vlined]{algorithm2e}
\usepackage {algpseudocode,comment}
\usepackage{algorithmicx}
\usepackage{algcompatible}
\input{mysymbol.sty}
\def\BibTeX{{\rm B\kern-.05em{\sc i\kern-.025em b}\kern-.08em
    T\kern-.1667em\lower.7ex\hbox{E}\kern-.125emX}}
\begin{document}

\title{Mitigation of Datacenter Demand Ramping and Fluctuation using Hybrid ESS and Supercapacitor
}

\author{\IEEEauthorblockN{Min-Seung~Ko and Hao~Zhu}
\IEEEauthorblockA{Chandra Family Department of Electrical \& Computer Engineering \\
The University of Texas at Austin\\
Austin, TX, USA \\
\{kms4634500, haozhu\}@utexas.edu}
\and
\IEEEauthorblockN{Jae~Woong~Shim}
\IEEEauthorblockA{Department of Electrical Engineering\\
Sangmyung University\\
Seoul, South Korea\\
jaewshim@smu.ac.kr}}

\maketitle

\begin{abstract}
This paper proposes a hybrid energy storage system (HESS)-based control framework that enables comprehensive power smoothing for hyperscale AI datacenters with large load variations. Datacenters impose severe ramping and fluctuation-induced stresses on the grid frequency and voltage stability. To mitigate such disturbances, the proposed HESS integrates a battery energy storage system (BESS) and a supercapacitor (SC) through coordinated multi-timescale control. A high-pass filter (HPF) separates the datacenter demand into slow and fast components, allocating them respectively to the ESS via a leaky-integral controller and to the SC via a phase-lead proportional-derivative controller enhanced with feedforward and ramp-tracking compensation. Adaptive weighting and repetitive control mechanisms further improve transient and periodic responses. Case studies verify that the proposed method effectively suppresses both ramping and fluctuations, stabilizes the system frequency, and maintains sustainable state-of-charge (SoC) trajectories for both ESS and SC under prolonged, stochastic training cycles.
\end{abstract}

\begin{IEEEkeywords}
Battery energy storage system, co-located load, datacenter, power smoothing, supercapacitor. 
\end{IEEEkeywords}

\section{Introduction} \label{Sec0}

Hyperscale AI datacenters have emerged as a critical load demand, driven by the rapid proliferation of large language models (LLMs) and other computationally intensive AI workloads \cite{nerc2025char}.
Although securing sufficient energization capacity remains an immediate concern, these power-hungry computing facilities have already led to significant challenges in grid reliability and stability. In particular, training LLM-typed large-scale AI models can require several hundred MW per site, and thus these workloads dominate the overall load patterns of hyperscale datacenters \cite{aljbour2024powering}. They are known to produce large and irregular power fluctuations, quickly becoming a major new threat to both local and grid-wide stability. Consequently, datacenter demand patterns are predominately characterized by abrupt ramping events and periodic fluctuations. Ramping behavior can cause severe local voltage stress, especially under weak-grid conditions. Meanwhile, periodic and sustained power fluctuations, with the characteristic frequency tied to the underlying large AI model training, can excite resonant oscillatory modes throughout the grid interconnection, as shown by our recent work\cite{ko2025wide}.


To enhance grid stability under increasing variability of datacenter loads, several mitigation strategies for power smoothing have been investigated. The rack level has mainly considered capacitive frequency-locking mechanisms, while energy storage systems (ESSs) are identified for the datacenter-wide mitigation. However, as demonstrated in \cite{choukse2025power}, these ESS-based solutions alone are insufficient to tackle the sharp transients induced by the start/end of AI workloads due to their inherent slower response timescales. Meanwhile, strengthening the grid infrastructure is also a viable approach, such as using grid-forming control of co-located storage to suppress sustained oscillations caused by fluctuating loads \cite{kundu2025managing}. Nevertheless, prior research has not fully addressed the physical limitations of standalone ESSs, and it remains a challenge to develop effective mitigation strategies that can simultaneously suppress the ramping and fluctuating demands of AI datacenters. 

To address this research gap, we propose a hybrid energy storage system (HESS) by co-locating a supercapacitor (SC) with an ESS to attain power smoothing across multiple timescales. Similar HESS solutions have been used to smooth intermittent renewable power generation, with applications in wind farms \cite{9138415} and solar plants \cite{xiao2015hierarchical}. Although these approaches are effective for slow or quasi-periodic variations, they largely rely on instantaneous power-error feedback and thus provide limited and insufficient treatment for large fast ramp events, typical of AI datacenter loads. 

Hence, this paper develops an adaptive HESS control framework that combines frequency-based power decomposition and a look-ahead compensation approach to systematically mitigate load ramps in addition to fluctuations. We use the SC to tackle high-frequency and sharp transients and the ESS to regulate slower variations and maintain the long-term energy balance. In parallel, a Kalman-filter (KF) observer is introduced to estimate the instantaneous slope and jerk in the demand profile, enabling a predictive feedforward that prepares the SC ahead of steep ramps and improves transient tracking beyond pure feedback control. The SC control loop employs a phase-lead proportional-derivative (PD) controller, while the ESS loop employs a leaky integral controller for sustained correction and suppression of low-frequency changes. A unified SoC-reference manager is further used to ensure energy sustainability with minimal deviation from the desired grid power reference. Collectively, these elements can effectively suppress both ramping and oscillatory transients for enhancing the stability of grid operations.

\section{System Modeling of AI Datacenters} \label{Sec1}

\begin{figure}[t]
    \centering
    \begin{tabular}{c}
    \includegraphics[width=0.93\columnwidth]{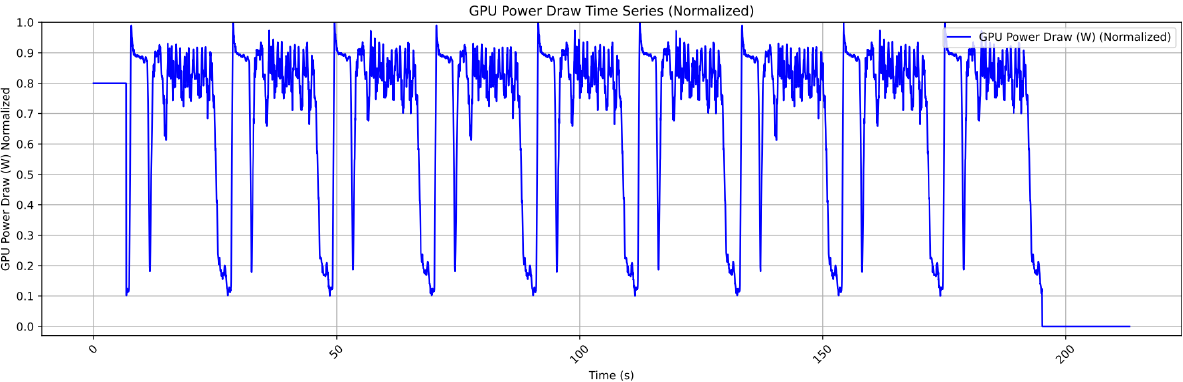} \\
    \small (a) \\
    \includegraphics[width=0.93\columnwidth]{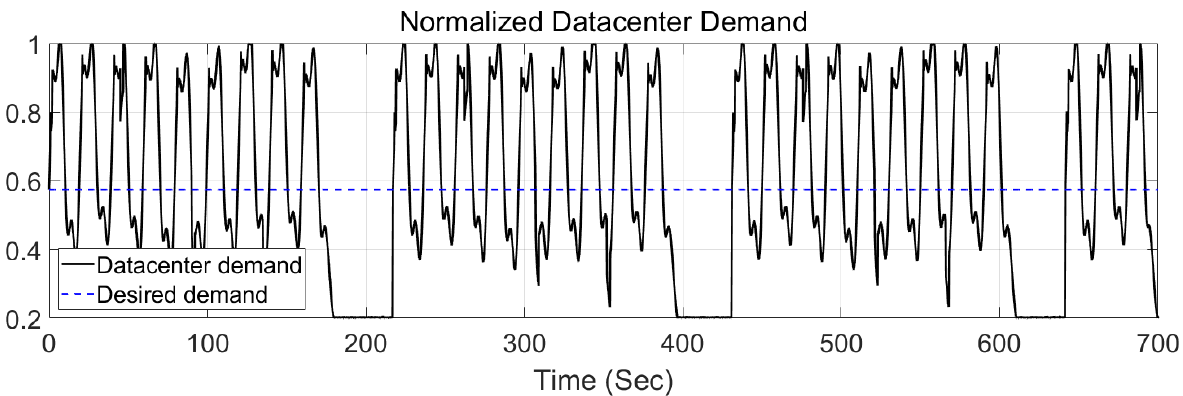} \\
    \small (b)    
    \end{tabular}
    \caption{Schematic diagrams of (a) realistic GPU power consumption from \cite{choukse2025power} and (b) synthetic datacenter demand.}
    \label{dcdemand}
\end{figure}

We first present the modeling of power fluctuations induced by AI training workloads in hyperscale AI datacenters. Fig.~\ref{dcdemand}(a) illustrates the example measurements of the GPU-level power trace, which exhibit both quasi-periodic fluctuations and abrupt ramping patterns. The fluctuation component is primarily driven by cyclic variations with approximately 0.05 Hz, arising from the iterative gradient computation inherent to the training process. Within these dominant cycles, higher-frequency sub-fluctuations and stochastic variations also exist, reflecting the fine-grained computational dynamics of GPUs. In terms of ramp behavior, the measured trace clearly demonstrates full-range transitions between idle and peak operating conditions due to the high GPU utilization.

Based on these measurement characteristics, we have synthetically generated the aggregate power demand of an AI datacenter. An example of the generated profile is illustrated in Fig.~\ref{dcdemand}(b). Unlike the GPU-level trace, the datacenter-level load has additional components beyond the server-computing loads. Moreover, in large-scale LLM training, multiple GPUs are synchronized by the training cycle to produce coherent load variations. Accordingly, the synthetic datacenter load demand retains the dominant fluctuation with some modifications to reflect the aggregated behavior. The sub-fluctuation component becomes more regular, resulting in smoother high-frequency oscillations. In addition, even after the training terminates, a non-zero baseline power is preserved to account for persistent non-computational loads. After aggregation, the two dynamic variations, namely periodic fluctuations and ramping patterns, are the most dominant in the total datacenter power profile. 

\begin{figure}[t]
    \centering
    \includegraphics[width=0.9\columnwidth]{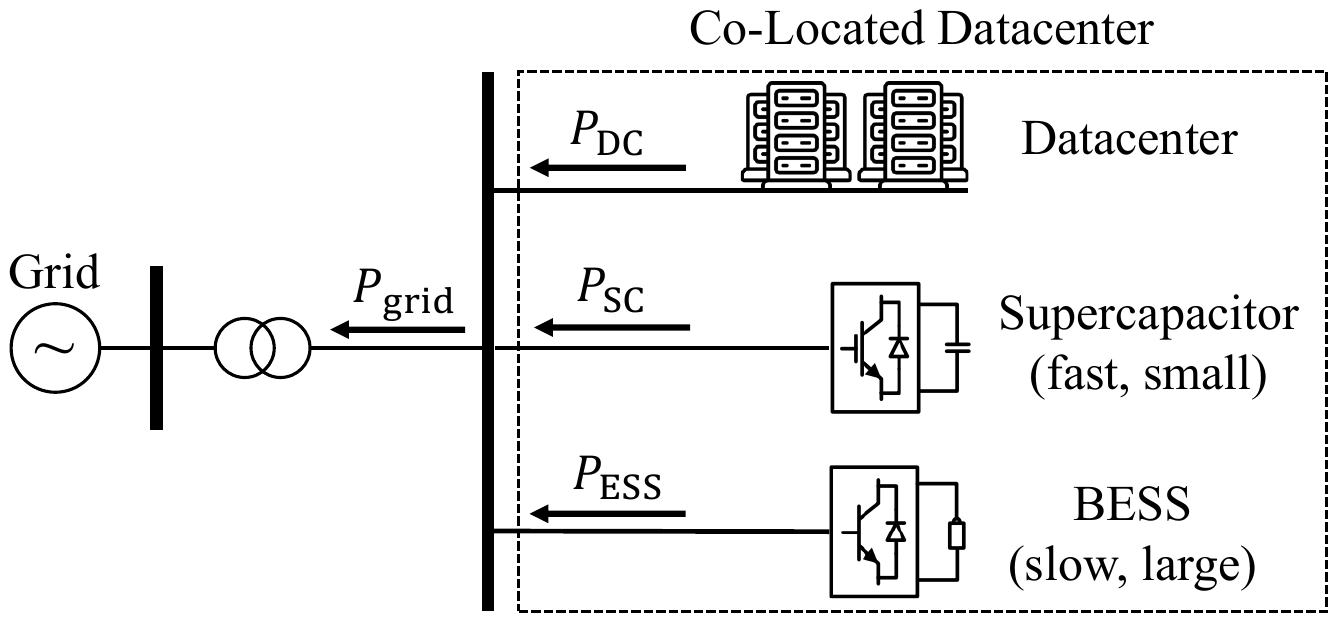}
    \caption{Configuration of the co-located datacenter with HESS.}
    \label{config}
\end{figure}

To address both dynamic variations simultaneously, we propose co-located hybrid resources composed of a battery ESS (BESS) and a supercapacitor (SC) in the AI datacenter. The proposed hybrid energy storage system (HESS) is connected to the same point of interconnection (POI) in parallel with the datacenter, as illustrated in Fig.~\ref{config}. The total power supply by the grid at time $t$, $P_\mathrm{grid}(t)$, can be represented as
\begin{align}
    P_\mathrm{grid}(t)=P_\mathrm{DC}(t)+P_\mathrm{ESS}(t)+P_\mathrm{SC}(t)
    \label{pgrid}
\end{align}
where $P_\mathrm{DC}$ stands for the datacenter load. The outputs of BESS and SC, $P_\mathrm{ESS}$ and $P_\mathrm{SC}$, indicate the discharged power when they have positive values. The HESS is operated under the constraints on power, ramp, and state of charge (SoC), as given by $|P_{(\cdot)}|\leq P^{\mathrm{max}}_{(\cdot)}$, $|R_{(\cdot)}|\leq{R^{\mathrm{max}}_{(\cdot)}}$, and $\mathrm{SoC}^{\mathrm{min}}_{(\cdot)}\leq\mathrm{SoC}_{(\cdot)}\leq\mathrm{SoC}^{\mathrm{max}}_{(\cdot)}$, respectively. Here, the subscript symbol $(\cdot)$ can represent either BESS or SC and $R$ stands for ramp rate. Based on the standard modeling approaches in \cite{9138415}, each device is represented by a first-order dynamic system in the following Laplace form in the $s$-domain:
\begin{align}
    P_{(\cdot)}(s)=G_{(\cdot)}(s)U_{(\cdot)}(s)=\frac{1}{\tau_{(\cdot)}s+1}U_{(\cdot)}(s)
    \label{fods}
\end{align}
with $U_{(\cdot)}(s)$ denoting the controller output for managing the corresponding power output $P_{(\cdot)}(s)$. 
Here, the parameter $\tau_{(\cdot)}$ stands for the given time constant related to the dynamic speed of each device. In particular, BESS tends to have a much larger capacity than SC, while the SC can provide a faster response to load deviation than BESS. Hence, the parameter settings should follow the below orderings:
\begin{align}
    P^{\mathrm{max}}_{\mathrm{ESS}} \gg P^{\mathrm{max}}_{\mathrm{SC}},~ \dot{P}^{\mathrm{max}}_{\mathrm{ESS}} \ll \dot{P}^{\mathrm{max}}_{\mathrm{SC}},~\textrm{and}~\tau_{\mathrm{ESS}} \gg \tau_{\mathrm{SC}}.
\end{align}

\section{Power Smoothing via HESS Control Designs} \label{Sec2}

\begin{figure}[t]
    \centering
    \begin{tabular}{c}
    \includegraphics[width=0.95\columnwidth]{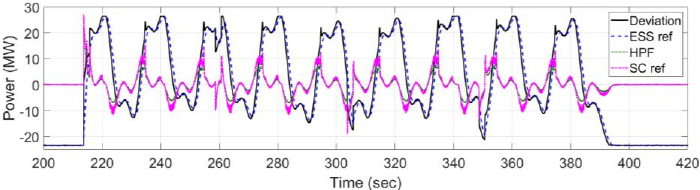} \\
    \small (a) \\
    \includegraphics[width=0.95\columnwidth]{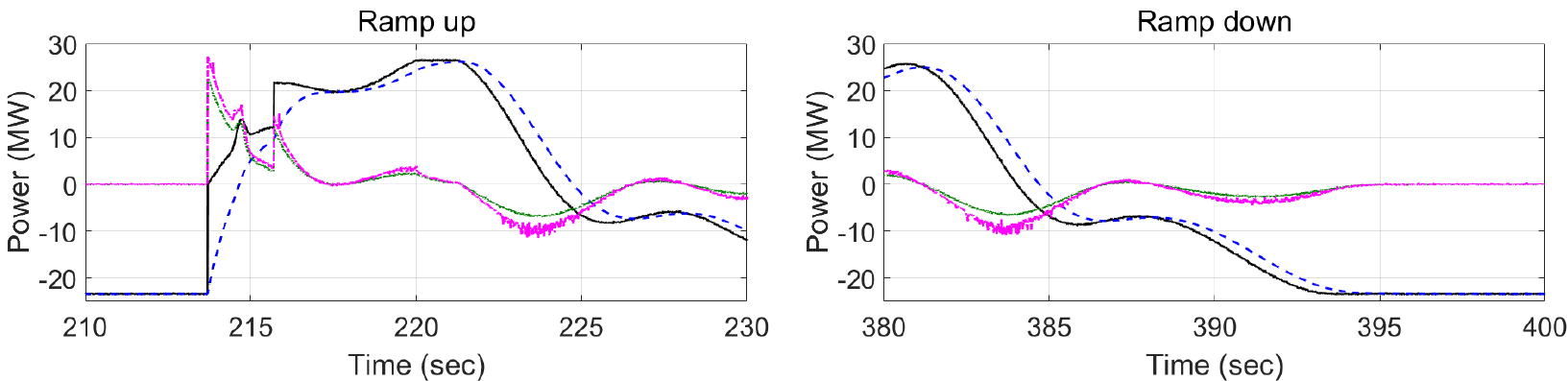} \\
    \small (b)
    \end{tabular}
    \caption{Command signals (a) in overall timeframe and (b) during ramp.}
    \label{refsig}
\end{figure}

The goal of proposing to co-locate HESS in datacenters is to attain power smoothing of $P_{\mathrm{DC}}(t)$ by utilizing the two heterogeneous and complementary resources. As discussed in Section~\ref{Sec1}, SC has a much faster response than BESS, motivating us to design power command signals of different frequency ranges that can match the two resources. We first define $\Delta(t)=P_{\mathrm{DC}}(t)-P_{\mathrm{DC}}^0$ as the deviation from its DC frequency component $P_{\mathrm{DC}}^0$. Thus, the SC power command uses the higher-frequency component of $\Delta(t)$, while the remaining is assigned to the BESS, as given by
\begin{align}
    P_{\mathrm{SC}}^{\mathrm{ref}0}&=G_{\mathrm{HPF}}(s)\Delta(s)=\frac{s}{s+2{\pi}f_{\mathrm{c}}}\Delta(s) \label{command1}\\
    P_{\mathrm{ESS}}^{\mathrm{ref}0}&=\left(1-G_{\mathrm{HPF}}(s)\right)\Delta(s), \label{command1_2}   
\end{align}
where $f_{\mathrm{c}}$ denotes the cutoff frequency for the high-pass filtering (HPF). This way, the SC responds to the rapid variations and transient spikes, while the BESS compensates for slower deviations by using a larger energy capacity than SC. Examples of the filtered signals are illustrated in Fig.~\ref{refsig}.

In the event of severe ramping or sharp fluctuations, the SC with the HPF signal would still exhibit some relative response delay that hinders a timely response. To address this issue, we incorporate an additional ramp-related term into the SC control loop, as
\begin{align}
    P_{\mathrm{SC}}^{\mathrm{ref}1}=G_{\mathrm{HPF}}(s)\Delta(s)+\omega_{\mathrm{ramp}}T_{\mathrm{eff}}\hat{R}_{\mathrm{DC}}(s),
    \label{command2}
\end{align}
where $\hat{R}_{\mathrm{DC}}(s)$ is the estimated ramp rate with an adaptive scaling factor $\omega_{\mathrm{ramp}}$, which will be discussed soon. In addition, the look-ahead horizon, $T_{\mathrm{eff}}$ is selected to be close to $\tau_\mathrm{SC}$, the SC's time constant, for this command signal to lead a single time step. As illustrated in Fig.~\ref{refsig}(b), this enhanced SC command design can provide more adequate reference values for ramp conditions.

To estimate the ramp rate $\hat{R}_{\mathrm{DC}}$ in \eqref{command2}, we employ a discrete-time Kalman filter (KF). For a discrete timestep $k$ with a sampling interval $T_s$, the KF state vector is defined as $\mathbf{z}^{\mathrm{KF}}_k=\begin{bmatrix} \bar{\Delta}, \; R_\mathrm{DC} \end{bmatrix}^\top$. Here, $\bar{\Delta}$ represents a slow-varying baseline of the deviation signal $\Delta(t)$ and its rate of change defines the ramp rate $R_\mathrm{DC}$. Using a time-varying baseline allows the KF to capture the slow drift handled by the ESS, so that $\hat{R}_{\mathrm{DC}}$ reflects the actual fast component that the SC must compensate for. Then, the KF relies on a second-order stochastic model expressed as 
\begin{align}
    \mathbf{z}^{\mathrm{KF}}_{k+1}=
    \begin{bmatrix}
        1 & T_s \\ 0 & \phi    
    \end{bmatrix}\mathbf{z}^{\mathrm{KF}}_{k}+\mathbf{w}_{k}, \quad
    \Delta_{k}=\begin{bmatrix}
        1 & 0    
    \end{bmatrix}\mathbf{z}^{\mathrm{KF}}_{k}+v_k.
\end{align}
Here, the parameter $\phi$ is chosen to have a very slow ramp dynamics, while $\mathbf{w}_k$ and $v_k$ denote the process and measurement noises, respectively. By using the nonlinear KF estimation approach in \cite{orderud2005comparison}, one can update both states recursively to obtain $\hat{\bar{\Delta}}$ and  $\hat{R}_{\mathrm{DC}}$.
In addition, we design an adaptive update for the parameter $\omega_{\mathrm{ramp}}=\gamma_{\mathrm{ramp}}\cdot\gamma_{\mathrm{jerk}}$ by forming two scalar variables related to ramp and jerk, respectively, as given by
\begin{align}
    \gamma_{\mathrm{ramp}}=\textrm{Sigmoid}\left(\frac{|\hat{R}_{\mathrm{DC}}|}{s^{\mathrm{ref}}}\right)\!,~\gamma_{\mathrm{jerk}}= \left(1+\frac{|\hat{R}_{\mathrm{DC}}|}{A^{\mathrm{ref}}}\right)^{-1}\!. \label{gamr}
\end{align}
In \eqref{gamr}, the $\textrm{Sigmoid}$ function uses the logistic function, while the two parameters $s^\mathrm{ref}$ and $A^\mathrm{ref}$ are predefined by the reference slope and jerk, respectively. This way, the ramp term $\gamma_{\mathrm{ramp}}$ is activated only when the estimated ramp exceeds $s^\mathrm{ref}$. In addition, the value of $\gamma_\mathrm{jerk}$ is reduced during rapidly changing ramps, thereby mitigating the potential overshoot in the control response.

Lastly, a dynamic SoC command manager is introduced to ensure the long-term SoC balance of both BESS and SC. A slowly varying offset $q_{(\cdot)}^{\mathrm{ref}}$ biases the BESS and SC commands according to their SoC deviation $E_{(\cdot)}^\mathrm{SoC}(s)=\mathrm{SoC}_{(\cdot)}^{*}-\mathrm{SoC}_{(\cdot)}$. The offset evolves as
\begin{align}
    q_{(\cdot)}^{\mathrm{ref}}(s)=-\frac{k_{q,(\cdot)}}{s+1/T_q}E_{(\cdot)}^\mathrm{SoC}(s),    
\end{align}
and decays exponentially within a small deadband. Here, $k_q$ and $T_q$ are the gain and time constant of the bias filter, respectively. The final power commands are adjusted as
\begin{align}   P_{\mathrm{ESS}}^{\mathrm{ref}}=P_{\mathrm{ESS}}^{\mathrm{ref}0}+q_{\mathrm{ESS}}^{\mathrm{ref}},\quad P_{\mathrm{SC}}^{\mathrm{ref}}=P_{\mathrm{SC}}^{\mathrm{ref}1}+q_{\mathrm{SC}}^{\mathrm{ref}}.\label{finalref}
\end{align} 
This slow biasing loop maintains sustainable SoC trajectories and prevents long-term drift without disturbing the fast HPF-based power control.

\subsection{Design of BESS and SC Controllers}
To accurately track the command signals given in \eqref{finalref}, we will design the controller outputs for BESS and SC, namely $U_{\mathrm{ESS}}$ and $U_{\mathrm{SC}}$ [cf.~\eqref{fods}]. 
Due to SC's fast timescales, it adopts a proportional–derivative (PD) controller to provide phase lead compensation, while BESS employs an integral controller to address slow variations and maintain the long-term energy balance. This complementary design allows the HESS to achieve coordinated control across multiple frequency bands.

To improve tracking ability during the ramp, the enhanced SC control structure consists of three components: PD controller, feedforward (FF), and ramp tracking (RK). Using the SC power tracking error $\epsilon_{\mathrm{SC}}=P_{\mathrm{SC}}^{\mathrm{ref}}-P_{\mathrm{SC}}$, one sets the SC controller to be
\begin{align}
U_{\mathrm{SC}}(s)=C_{\mathrm{PD}}(s)\epsilon_{\mathrm{SC}}(s)+U_{\mathrm{FF,SC}}(s)+U_{\mathrm{RK}}(s).
\end{align}
We discuss each of the three terms here. First, the derivative action within the PD control can amplify high-frequency noise if left unregulated. To mitigate this effect, a first-order derivative filter is used to find 
\begin{align}
    C_{\mathrm{PD}}(s)=G_{\mathrm{HPF}}(s)\left(k_{p}+k_{d}\frac{2{\pi}f_{d}s}{s+2{\pi}f_{d}}\right)
\end{align}
with the derivative cutoff frequency $f_d$. Note that the PD controller has incorporated $G_{\mathrm{HPF}}(s)$ in \eqref{command1} to focus on the high-frequency range. Second, although the ramp component is implicitly included by the PD structure, a pure feedback control may still lag during abrupt, sharp transients. Therefore, we incorporate $U_{\mathrm{FF,SC}}(s)$ to provide partial delay compensation and improve the match with the SC reference at the beginning phase, given by
\begin{align}
    U_{\mathrm{FF,SC}}(s)=\left(s+\frac{1}{\tau_{\mathrm{SC}}}\right)\left(P_{\mathrm{SC}}^{\mathrm{ref}}(s)-\frac{P_{\mathrm{SC}}(s)}{e^{sT_s}}\right). \label{ffsc}
\end{align}
By reducing the burden on the feedback loop and mitigating phase lags inherent in the converter dynamics, this FF term helps to suppress overshoot while preserving adequate phase margin. Last, the RK term is used to address any remaining mismatch between the desired and actual ramp profiles, as
\begin{align}
    U_{\mathrm{RK}}(s)=k_{\mathrm{RK}}\left(\hat{R}_{\mathrm{DC}}(s)-\hat{R}_{\mathrm{SC}}(s)\right)
\end{align}
with the coefficient $k_\mathrm{RK}$ determining the correction gain. This compensator injects a corrective action based on the slope error between the commanded and measured power trajectories. To recap, the PD term responds to instantaneous errors and the FF term mitigates delay-induced mismatch, while the RK term ensures accurate tracking of sharp ramp segments, especially when the SC output lags behind the desired trajectory.

Similarly to SC, the BESS controller also consists of three components, the integral (I) controller, the FF loop, and the repetitive controller (RC), as expressed by
\begin{align}
    U_{\mathrm{ESS}}(s)=\left(C_{\mathrm{I}}(s)+C_{\mathrm{RC}}(s)\right)\epsilon_{\mathrm{ESS}}(s)+U_{\mathrm{FF,ESS}}(s).
\end{align}
First, the main feedback controller is implemented as a leaky I controller to prevent the windup issue:
\begin{align}
    C_{\mathrm{I}}(s)=\frac{1}{s+1/T_\mathrm{leak}}k_{\mathrm{I}}
\end{align}
where $T_\mathrm{leak}$ sets the leakage time constant and $k_{\mathrm{I}}$ is the integral gain. This formulation ensures good steady-state accuracy while maintaining a bounded control effort even under long-term energy imbalance conditions. In practice, this leaky I control also contributes to the energy-neutral operation of BESS by gradually restoring its SoC to the mid-level target. Second, the RC filter is introduced to suppress quasi-periodic deviations in the BESS tracking error that may persist due to the coupling with SC. The discrete comb filter \cite{chmielewski2021modified} is selected, which is approximated in the s-domain by
\begin{align}
    C_{\mathrm{RC}}(s)=k_{\mathrm{RC}}\frac{\omega_{\mathrm{RC}}}{s+\omega_{\mathrm{RC}}}
\end{align}
where $k_{\mathrm{RC}}$ is the compensation gain and $\omega_{\mathrm{RC}}$ corresponds to the dominant periodic frequency. This RC filter actively rejects repeating frequency components, thereby mitigating interaction-induced oscillations between the SC and BESS. Finally, the FF term for BESS is formulated with the same structure as in \eqref{ffsc} with different parameters, in order to reduce any delay-induced mismatch. 

\section{Experimental Settings and Results}

\begin{figure}[t]
    \centering
    \begin{tabular}{c}
    \includegraphics[width=0.95\columnwidth]{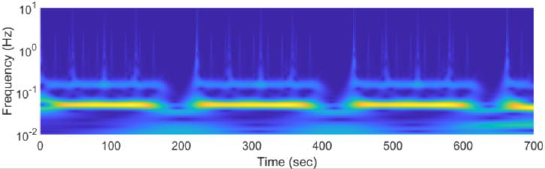} \\
    \small (a) \\
    \includegraphics[width=0.95\columnwidth]{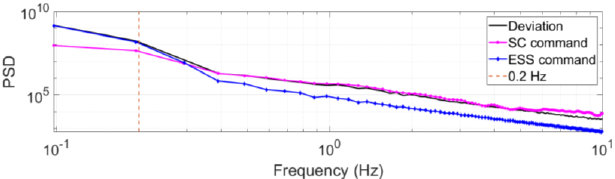} \\
    \small (b)
    \end{tabular}
    \caption{Schematic diagrams of (a) wavelet analysis results of $\Delta(t)$ and (b) power spectral density analysis results of command signals}
    \label{wavpsd}
\end{figure}

To investigate the impact of power smoothing through the HESS, we conduct case studies on the synthetic simplified 1 GW system. For simplicity, we assume a single-machine grid based on the swing equation and first-order governor dynamics. The deviation of datacenter demand is assumed to be 50 MW.
Detailed parameter values of the grid and HESS adopted for the experiment are shown in Table~\ref{simset}.
Specifically, we set the charge/discharge efficiency $\eta$ identical for both processes, as well as for BESS and SC.
To ensure realistic behavior, HESS controllers are implemented using the discrete-time formulation with $T_{s}=10$ ms. Furthermore, $\Delta(t)$ shows large energy near 0.05 Hz and 0.15 Hz, as shown in Fig.~\ref{wavpsd}(a). Notice that these frequencies correspond to the dominant and sub-periodic variations in Fig.~\ref{refsig}(a), respectively. To ensure that BESS can handle these large oscillatory components, we set the cutoff frequency at $f_{c}=0.2$ Hz.

\begin{table}[t]
\centering
\caption{Example of the Grid and HESS parameters}
\label{simset}
\begin{tabular}{c|c||c|c|c}
\hline
\multicolumn{2}{c||}{Grid Parameters} & HESS Parameters               & BESS     & SC \\ \hline
$H$                 & 6               & $P_\mathrm{max}$  (MW)        & 30   & 10       \\ \hline
$D$                 & 3               & $\dot{P}_\mathrm{max}$ (MW/s) & 15 & 100  \\ \hline
$R_\mathrm{droop}$  & 0.05            & $\tau$                        & 0.25    & 0.02 \\ \hline
$T_g$               & 0.3             & $\eta$                        & \multicolumn{2}{c}{0.97} \\ \hline
\end{tabular}
\begin{flushleft}{\footnotesize $H$: Inertia constant, $D$: Frequency-dependent damping, $R_{\mathrm{droop}}$: Governor droop rate, $T_{g}$: Governor time constant, $\eta$: Charge/discharge efficiency}\end{flushleft}
\end{table}

\begin{figure}[t]
    \centering
    \includegraphics[width=0.93\columnwidth]{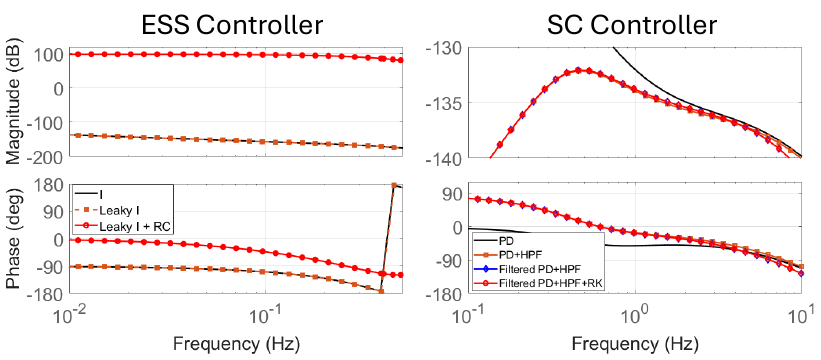} \\
    \vspace{-0.15cm}
    \caption{Open-loop bode plots of BESS and SC controllers.}
    \label{bode}
\end{figure}

To examine the frequency-domain characteristics of the designed command signals, a power spectral density (PSD) analysis is conducted, as shown in Fig.~\ref{wavpsd}(b). The PSD of $\Delta(t)$ exhibits a dominant concentration of energy at very low frequencies. However, a non-negligible portion remains at higher frequencies, confirming the need for multi-band control. With the chosen $f_c$, the BESS primarily covers frequencies below approximately 0.2 Hz, while SC responds dominantly to components above 0.4 Hz. This frequency separation enables the HESS to handle distinct dynamic regimes of the datacenter load. The open-loop Bode plots in Fig.~\ref{bode} further illustrate these complementary roles. For the SC loop, the transition from the basic PD to the proposed configuration progressively suppresses the low-frequency gain while maintaining an adequate mid-high frequency response; HPF and derivative filters also improve phase behavior and enhance robustness. In contrast, the BESS loop maintains a dominant low-frequency response, where the leaky I control prevents the drift, and the RC selectively boosts the gain near 0.05 Hz for improved rejection on periodic disturbance.

\begin{figure}[t]
    \centering
    \begin{tabular}{c}
    \includegraphics[width=0.95\columnwidth]{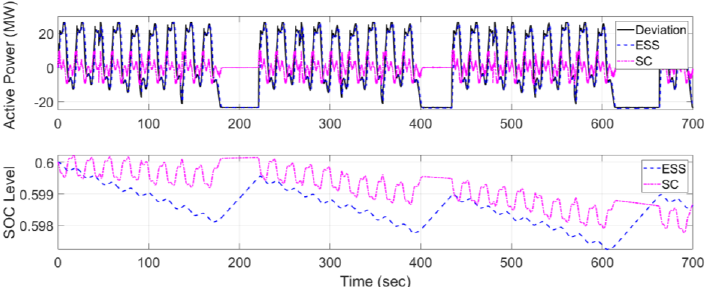} \\
    \small (a) \\
    \includegraphics[width=0.95\columnwidth]{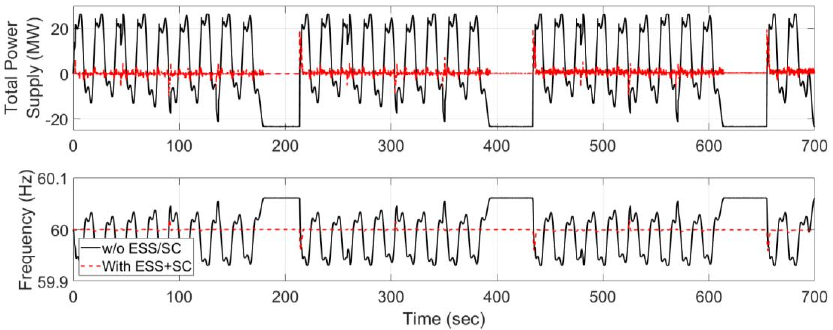} \\
    \small (b)    
    \end{tabular}
    \caption{Schematic diagrams of (a) power output and SoC level, and (b) total supplied power and frequency.}
    \label{result}
\end{figure}   

\begin{figure}[t]
    \centering
    \includegraphics[width=0.95\columnwidth]{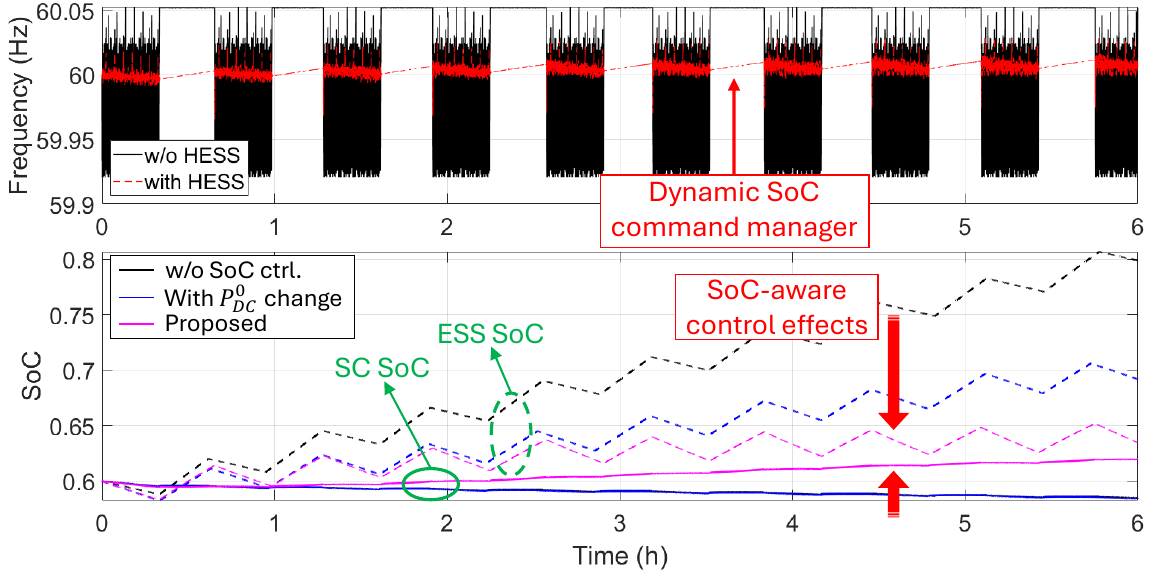} 
    \vspace{-0.1cm}
    \caption{Long-term simulation results with dynamic SoC command manager.}
    \label{long}
\end{figure}   

The resulting responses of the BESS and SC outputs to their respective command signals are depicted in Fig.~\ref{result}(a), while Fig.~\ref{result}(b) presents the total power supply and grid frequency with and without the HESS. Both BESS and SC closely follow their command trajectories, maintaining the grid-supplied power nearly constant and minimizing frequency deviations. During ramp-up events, a slight mismatch between the SC command and its actual output can be observed due to its physical power limit. This constraint leads to a transient increase in grid-supplied power, causing a momentary frequency dip. Nonetheless, these transient deviations are brief and remain within acceptable operating limits. In addition, although we do not implement the SoC manager in this experiment, SoCs of both components remain highly stable throughout the 10 min simulation, exhibiting less than 0.03\% variation. The BESS effectively recharges during idle periods, restoring its SoC toward the nominal operating level. However, both devices exhibit a gradual downward trend over longer horizons, suggesting potential energy imbalance during extended operation or prolonged training cycles.

To further evaluate this long-term performance, we conduct an additional simulation using a more realistic datacenter training profile characterized by longer active and idle periods. The corresponding results are shown in Fig.~\ref{long}. As observed, the SoC trajectories exhibit a gradual drift toward the upper limit, as a result of the sustained positive net energy injection during extended training phases. To mitigate this imbalance, we incorporate baseline correction of the desired demand and apply a dynamic SoC command manager during idle intervals. Notice that the frequency during the idle period experiences small deviations because of the change in reference from the SoC manager. With these enhancements, both components maintain their SoC within a narrow range of approximately 0.05\%, while the system frequency remains tightly regulated around the nominal 60 Hz. These results confirm that the proposed scheme ensures sustainable operation and long-term stability even under prolonged and stochastic load variations.

\section{Conclusion}
This work presents a coordinated hybrid control framework for power smoothing of AI datacenter loads using co-located BESS and SC units. By decomposing the demand signal through HPF, the controller effectively distributes control actions across complementary timescales: the BESS manages slow fluctuations and energy balancing, while the SC rapidly compensates for transient ramps. The inclusion of feedforward delay compensation, ramp-tracking, and repetitive control terms further enhances dynamic performance without violating power or SoC constraints. Simulation results confirm that the proposed HESS substantially reduces power and frequency deviations compared with standalone operation, while maintaining device sustainability during long-duration training. 

\ifCLASSOPTIONcaptionsoff
  \newpage
\fi
\bibliographystyle{IEEEtran}
\bibliography{ref}

\end{document}